\documentclass[letterpaper]{article}

\usepackage{proceed2e}
\usepackage{times,cuted}
\usepackage{epsfig}
\usepackage{graphicx}
\usepackage{amsmath}
\usepackage{amssymb}
\usepackage{appendix}

\usepackage[pagebackref=true,breaklinks=true,letterpaper=true,colorlinks,bookmarks=false]{hyperref}

\usepackage[accepted]{icml2017}
 
 \icmltitlerunning{Approximate Bayesian inference as a gauge theory}
\begin{document}


\twocolumn[
\icmltitle{Approximate Bayesian inference as a gauge theory}


\icmlsetsymbol{equal}{*}

\begin{icmlauthorlist}
\icmlauthor{Biswa Sengupta}{imp,cam,cor}
\icmlauthor{Karl Friston}{fil}
\end{icmlauthorlist}

\icmlaffiliation{imp}{Dept. of Bioengineering, Imperial College London}
\icmlaffiliation{cam}{Dept. of Engineering, University of Cambridge}
\icmlaffiliation{cor}{Cortexica Vision Systems, UK}
\icmlaffiliation{fil}{Wellcome Trust Centre for Neuroimaging, University College London}

\icmlcorrespondingauthor{Biswa Sengupta}{b.sengupta@imperial.ac.uk}

\icmlkeywords{Riemann geometry, gauge theory, Bayesian statistics}

\vskip 0.3in
]

\printAffiliationsAndNotice{} 

\begin{abstract}
In a published paper \cite{Sengupta2016}, we have proposed that the brain (and other self-organized biological and artificial systems) can be characterized via the mathematical apparatus of a gauge theory. The picture that emerges from this approach suggests that any biological system (from a neuron to an organism) can be cast as resolving uncertainty about its external milieu, either by changing its internal states or its relationship to the environment. Using formal arguments, we have shown that a gauge theory for neuronal dynamics -- based on approximate Bayesian inference -- has the potential to shed new light on phenomena that have thus far eluded a formal description, such as attention and the link between action and perception. Here, we describe the technical apparatus that enables such a variational inference on manifolds. Particularly, the novel contribution of this paper is an algorithm that utlizes a Schild's ladder for parallel transport of sufficient statistics (means, covariances, etc.) on a statistical manifold.
\end{abstract}

\section{Introduction}

A gauge theory is a physical theory that predicts how one or more physical fields interact with matter. Every gauge theory has an associated Lagrangian (i.e., a function that summarizes the dynamics of the system in question), which is invariant under a group of local transformations.  Consider Newton's laws of motion in an inertial frame of reference (e.g. a ball in an empty room). These laws are valid at every point in the room. This means that the dynamics and the Lagrangian do not depend on the position of the ball. In this system, the dynamics will be invariant under translations in space. These transformations -- that preserve the Lagrangian -- are said to be equipped with gauge symmetry. In short, a symmetry is simply an invariance or immunity to changes in the frame of reference. 

Gauge theories originate from physics; however, they could be applied to countless fields of biology: cell structure, morphogenesis, and so on. Examples that lend themselves to a gauge theoretic treatment include recent simulations of embryogenesis \cite{Friston2015} and the self-organisation of dendritic spines \cite{Kiebel2011}. Although these examples appear unrelated, both can be formulated in terms of a gradient ascent on variational free energy. In other words, we may be looking at the same fundamental behaviour in different contexts. Here, we focus on the central nervous system (CNS).  Can we sketch a gauge theory of brain function?  

When attempting to establish what aspect of CNS function might be understood in terms of a Lagrangian, the variational free energy looks highly plausible \cite{Friston2010}.  The basic idea is that any self-organizing system, at non-equilibrium steady-state with its environment, will appear to minimize its (variational) free energy, thus resisting a natural tendency to disorder.  This formulation reduces the physiology of biological systems to their homeostasis (and allostasis); namely, the maintenance of their states and form, in the face of a constantly changing environment.  

If the minimisation of variational free energy is a ubiquitous aspect of biological systems could it be the Lagrangian of a gauge theory? This (free energy) Lagrangian has proved useful in understanding many aspects of functional brain architectures; for example, its hierarchical organisation and the asymmetries between forward and backward connections in cortical hierarchies.  In this setting, the system stands for the brain (with neuronal or internal states), while the environment (with external states) is equipped with continuous forces and produces local sensory perturbations that are countered through action and perception (that are functionals of the gauge field). 

In summary, the free energy formalism rests on a statistical separation between the agent (the internal states) and the environment (the external states). Agents suppress free energy (or surprise) by changing sensory input, by acting on external states, or by modifying their internal states through perception. In what follows, we show that the need to minimize variational free energy (and hence achieve homeostasis) acquires a useful logical-mathematical formalism, when framed as a gauge theory.

\section{Methods}

\subsection*{Variational free-energy formalism}

The variational free energy formalism assumes that an agent minimizes the entropy of its sensory states $s \in S$. Only through its sensory receptors can a biological system access the states of its environment; in other words, sensory states form a veil (technically, a Markov blanket) between the system's internal states $\theta  \in \Theta $  and its environment (external states) $\psi  \in \Psi $. By bounding the entropy of its sensory states, the system confines the entropy of its environment. Under the assumption of ergodicity, this entropy is the long-term average of surprise. Crucially, the system cannot calculate this quantity directly because it has to marginalize over the external states that cause sensory input. The objective then becomes to obtain a lower bound on the marginal likelihood $\ln p(s)$  by approximating it using a parametric probability distribution (for example, a Gaussian distribution)   $q(\psi )$ over the (unknown) external states that are hidden behind the Markov blanket (i.e., hidden causes of sensations). In short, the Lagrangian $\mathcal{L} =  - \ln p(s)$  is minimised by bounding the surprise using the variational free energy $\mathcal{F}(s,\theta ) \triangleq \mathcal{F}(s,q)$  of the distribution $q(\psi |\theta )$  where $\theta $  are the sufficient statistics or parameters of the variational distribution. In the case of a Gaussian distribution the sufficient statistics are simply the mean and the co-variances  $\{ \mu ,\sum \}  \subset \theta $. 

The variational distribution that minimises free energy can be expressed in terms of an Euler-Lagrange action  $\delta S(\mathcal{F}(s,q)) = 0 \Leftrightarrow \nabla \mathcal{F} = 0$, implying that the gradient descent on the variational free-energy manifold leads us to the most optimal representation of the external states.  A numerical scheme to solve such a variational problem is generalised (Bayesian) filtering \cite{Friston2008}. The Bayesian perspective follows because our Lagrangian is also known as (the negative logarithm of) Bayesian model evidence. In other words, minimising free energy is equivalent to maximising model evidence.\\

\subsection*{Variational inference on manifolds}

The evolution of the variational free-energy could be described on a Riemann manifold by augmenting the first order gradient flow using a Fisher information metric \cite{Amari1995,Tanaka2001}.  On a Euclidean manifold, the minimization of variational free-energy involves

\begin{eqnarray}
\frac{{ - \nabla \mathcal{F}}}{{\left\| {\nabla \mathcal{F}} \right\|}} = \mathop {\lim }\limits_{\varepsilon  \to 0} \frac{1}{\varepsilon }\mathop {\arg \min }\limits_{d\theta :\left\| {d\theta } \right\| \leqslant \varepsilon } \mathcal{F}(\theta  + d\theta )
  \label{eqn:euclidean}
\end{eqnarray}

This simply says that the flow of parameters (e.g. means and co-variance of a Normal distribution; $\{ \mu ,\sigma , \ldots \}  \in \theta $) will induce the largest change in free-energy under a unit change in parameters. Notice that the inner products are defined on a Euclidean manifold. 

Classical results from information geometry (Cencov's characterisation theorem) tell us that, for manifolds based on probability measures, a unique Riemannian metric exists -- the Fisher information metric. In statistics, Fisher-information is used to measure the expected value of the observed information. Whilst the Fisher-information becomes the metric for curved probability spaces, the distance between two distributions is provided by the Kullback-Leibler (KL) divergence. It turns out that if the KL-divergence is viewed as a curve on a curved surface, the Fisher-information becomes its curvature:

\begin{strip}
\begin{eqnarray}
  K{L_{sym}}(\theta ,\theta ') & = & {{\rm E}_\theta }\left[ {\log \frac{{q(\psi \left| {\theta )} \right.}}{{q(\psi \left| {\theta ')} \right.}}} \right] + {{\rm E}_{\theta '}}\left[ {\log \frac{{q(\psi \left| {\theta ')} \right.}}{{q(\psi \left| {\theta )} \right.}}} \right] \nonumber \\ 
   & = & d{\theta ^T}g(\theta )d\theta  + \mathcal{O}(d{\theta ^3}) \nonumber \\ 
  {g_{ij}}(\theta ) & = & \int\limits_{ - \infty }^{ + \infty } {q(\psi ,\theta )} \frac{{\partial \ln q(\psi ,\theta )}}{{\partial {\theta _i}}}\frac{{\partial \ln q(\psi ,\theta )}}{{\partial {\theta _j}}}d\psi  \nonumber \\ 
  KL[{\theta _0} + \delta \theta :{\theta _0}] & \doteq & \frac{1}{2}{g_{ij}}({\theta _0}){(\delta \theta )^2} \nonumber \\
  \label{eqn:fisher}
\end{eqnarray}
\end{strip}
	 	
Gradient descent on such a manifold then becomes the solution of $\arg \mathop {\min }\limits_{d\theta } \mathcal{F}(\theta  + d\theta )$, subject to $K{L_{sym}}(\theta ,\theta  + d\theta ) < \varepsilon$; i.e., the direction of the highest decrease in the free-energy, for the smallest change in the KL divergence. The solution of this optimization problem yields Amari's natural gradient that replaces the Euclidean gradient  $\nabla \mathcal{F}$ by its Riemannian counterpart $\tilde \nabla \mathcal{F} = {{\text{g}}_{{\text{ij}}}}{{\text{(}}\theta {\text{)}}^{{\text{ - 1}}}}\nabla \mathcal{F}$. This derivative is invariant under re-parameterisation of the approximate probability distribution, thereby helping us to break symmetries on the variational free-energy manifold. 

This formulation has two important consequences -- (a) from classical results in statistics, pre-conditioning of the free-energy gradient by the Fisher-information tells us that the variance of the estimator is bounded from below by the Fisher-information (Cram\'{e}r-Rao bound) and (b) under a Normal distribution approximation of the posterior distribution, precision-weighted prediction errors under a Euclidean manifold are replaced by asymptotic dispersion and precision-weighted prediction errors under a Riemannian manifold.

Such constructs are already instantiated in advanced Bayesian filtering schemes, such as the Statistical Parametric Mapping (SPM) code-base (available from \url{http://www.fil.ion.ucl.ac.uk/spm/}) using Fisher-scoring -- the gradient of variational free-energy is pre-multiplied by the inverse Fisher information metric. Notice that the metric in Fisher-scoring is simply the variance of the score function, while our derivation of the metric includes not only the metric for the likelihood but also that of the prior (instantiated as the Hessian of the prior). 

\begin{figure}
 \centering \includegraphics[width=0.5\textwidth, height=1in]
 {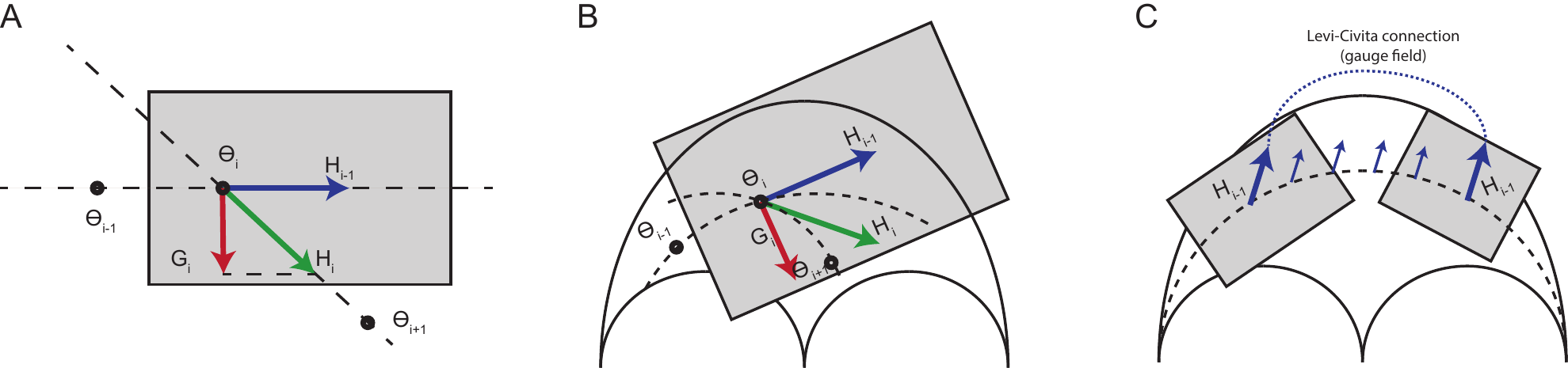}
 \caption{. Conjugate gradient-descent algorithm on manifolds. (A) New parameters $(\theta )$  are selected by performing gradient descent on orthogonal sub-spaces with gradient $G$  and the descent direction  $H$. (B) On a Riemannian manifold, minimization along lines (as in a Euclidean sub-space described in A) is replaced by minimization along geodesics. This creates a problem, in that ${H_i}$  and ${H_{i - 1}}$  are in two different tangent spaces and thus cannot be added together. (C) Vector addition as in Eqn. \ref{eqn:cg} is undefined on a Riemannian manifold. Addition is replaced by exponential mapping followed with parallel transport described using a co-variant gauge field (Levi-Civita connection; see text)}
 \label{fig_s1}
\end{figure}

The question that we now ask is whether we can deduce an optimization scheme that enables us to traverse the free-energy landscape? In other words, find the geodesic to local minima in the sub-manifold. There are two routes one can take to increase the statistical efficiency of the implicit optimisation -- first, we can formulate the Hessian operator on the Riemannian manifold in terms of the Laplace-Beltrami operator (Section S3.1 in \cite{Sengupta2016}) or we can retain a first-order approximation and formulate descent directions that are orthogonal to the previous descent directions. Such Krylov sub-spaces are well-known in numerical analysis with the conjugate gradient-descent algorithm providing one such example (Figure \ref{fig_s1}). Routinely used in optimization, conjugate gradient descent methods have been used for gradient descent on manifolds traced out by energy functions such as the variational free-energy \cite{Hensman,Honkela2010}. Simply such a scheme amounts to,

\begin{eqnarray}
  {\theta _i} & = & {\theta _{i - 1}} + \alpha {H_i} \nonumber \\ 
  {H_i} & = &  - {G_i} + \beta {H_{i - 1}} \nonumber \\
  \beta  & = & \frac{{\nabla _i^T\nabla _i^{}}}{{\nabla _{i - 1}^T\nabla _{i - 1}^{}}}\mathop  \to \limits^{Riemannize} \frac{{\tilde \nabla _i^T\nabla _i^{}}}{{\tilde \nabla _{i - 1}^T\nabla _{i - 1}^{}}} \nonumber \\ 
 \label{eqn:cg}
\end{eqnarray}
 	
For $\beta $ we have used the Fletecher-Reeves instantiation on a curved manifold; other update rules such as Polak-Ribi\`{e}re, Hestenes-Stiefel or Dai-Yuan can be similarly lifted to a Riemannian manifold by altering the implicit norm. All of these conjugate gradient descent formulas have a problem -- one cannot add two vector fields ${H_i}$  and ${H_{i - 1}}$ on a Riemannian manifold. This is because they exist on different tangent manifolds. ${H_{i - 1}}$ should undergo parallel transport to the tangent manifold containing ${H_i}$ using a connection (a gauge) field. In our case, this is the Levi-Civita connection described in Section S2 in \cite{Sengupta2016}.

\begin{figure}
 \centering \includegraphics[width=0.5\textwidth, height=1in]
 {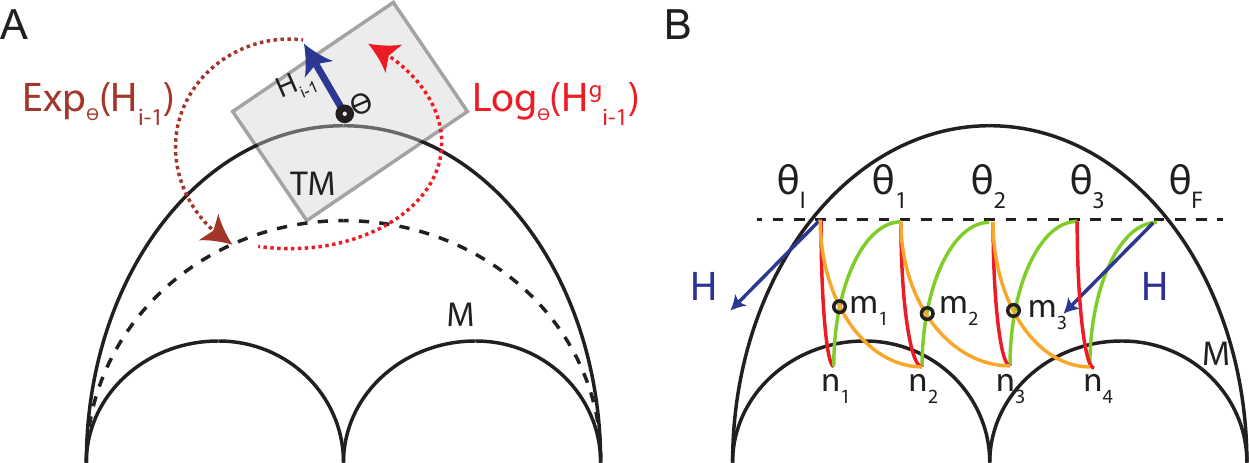}
 \caption{Parallel transport on Riemannian manifolds. (A) The Riemann exponential map is used to map a vector field $H$ from $T\mathcal{M} \to \mathcal{M}$ whilst a logarithmic map is used to map the vector field from $\mathcal{M} \to T\mathcal{M}$. (B) Graphical illustration of parallel transporting a vector field $H$ using a Schild's ladder (see text for details). }
 \label{fig_s2}
\end{figure}

Parallel-transport requires the solution of a second-order differential equation. Analysis shows us that the natural gradient is the first order approximation of the parallel transport -- that we pursue in terms of solving geodesic equations for the sufficient statistics. For the Laplace approximation, we could derive the Christoffel symbols analytically (Section S4 in \cite{Sengupta2016}), while for more complicated probability distributions we need to resort to a generic transport procedure (Figure \ref{fig_s2}). Namely, we use the Riemann exponential map for mapping the vector field on the tangent manifold to the geodesic described on the manifold $(T\mathcal{M} \to \mathcal{M})$, whereas a Riemann logarithmic map represents the transformations of vector-fields from the manifold to the tangent manifold $(\mathcal{M} \to T\mathcal{M})$,

\begin{eqnarray}
  {\exp _\theta }(A) = {\theta ^{1/2}}\exp ({\theta ^{ - 1/2}}A{\theta ^{ - 1/2}}){\theta ^{1/2}} \nonumber \\  
  {\log _\theta }(A) = {\theta ^{1/2}}\log ({\theta ^{ - 1/2}}A{\theta ^{ - 1/2}}){\theta ^{1/2}} \nonumber \\ 
 \label{eqn:exp}
\end{eqnarray}
	 	
The geodesic is first approximated using standard projection method \cite{Hairer2004}. Then using the exponential and logarithm maps a Schild's ladder \cite{Misner1973} is instantiated as following: Let  ${\theta _I}$ and ${\theta _F}$ denote the initial and final points on the geodesic that the vector field is to be transported to. We start by calculating ${n_1} = ex{p_{{\theta _I}}}(H)$ and the midpoint ${m_1}$ between the geodesic segment joining ${n_1}$ and ${\theta_1}$. We then trace out the geodesic from  ${\theta _I}$ through ${m_1}$ for twice its length, tracing out a new point ${n_2}$. This scheme is repeated until we reach ${\theta _F}$. After the vector field $H$ has been parallel transported to ${\theta _F}$, we are in a position to use parameter updates as detailed in Eqn. \ref{eqn:cg}. This scheme will be made available in upcoming versions of SPM for a variety of dynamical systems.

\section{Results}

\subsection*{Sensory entropy as a Lagrangian}

The variational free energy formalism uses the fact that biological systems must resist the second law of thermodynamics (i.e. a tendency to disorder), so that they do not decay to equilibrium. In a similar vein to Maxwell's demon, an organism reduces its entropy through sampling the environment -- to actively minimise the self information or surprise of each successive sensory sample (this surprise is upper bounded by free energy). By doing so, it places a bound on the entropy of attributes of the environment in which it is immersed. Variational free energy operationalises this bound by ensuring internal states of the system become a replica (i.e., a generative model) of its immediate environment. This can be regarded as a formulation of the good regulator hypothesis, which states that every good regulator of a system must be a model of that system. 

We know that a gauge theory would leave the Lagrangian invariant under continuous symmetry transformations. Therefore, a gauge theory of the brain requires the interaction among three ingredients: a system equipped with symmetry, some local forces applied to the system and one or more gauge fields to compensate for the local perturbations that are introduced.  The first ingredient is a system equipped with symmetry: for the purposes of our argument, the system is the nervous system and the Lagrangian is the entropy of sensory samples (which is upper-bounded by variational free energy, averaged over time). The local forces are mediated by the external states of the world (i.e., through sensory stimuli). The gauge fields can then be identified by considering the fact that variational free-energy is a scalar quantity based on probability measures.  

How does neuronal activity follow the steepest descent direction to attain its free energy minimum? In other words, how does it find the shortest path to the nearest minimum? As the free energy manifold is curved there are no orthonormal linear co-ordinates to describe it. This means the distance between two points on the manifold can only be determined with the help of the Fisher information metric that accounts for the curvature. Algebraic derivations (S3 in \cite{Sengupta2016}) tell us that, in such free-energy landscapes, a Euclidean gradient descent is replaced by a Riemann gradient, which simply weights the Euclidean gradient by its asymptotic variance. 

In the free energy framework, when the posterior probability is approximated with a Gaussian distribution (the Laplace approximation; S4 in \cite{Sengupta2016}), perception and action simply become gradient flows driven by precision-weighted prediction errors. Here, prediction errors are simply the difference between sensory input (local perturbations) and predictions of those inputs based upon the systems internal states (that encode probability distributions or Bayesian beliefs about external states that cause sensory input). Mathematically, precision-weighted prediction errors emerge when one computes the Euclidean gradient of the free energy with respect to the sufficient statistics. In a curvilinear space, the precision-weighted prediction errors are replaced by dispersion and precision weighted prediction errors.  This says something quite fundamental -- perception cannot be any more optimal than the asymptotic dispersion (inverse Fisher information) regardless of the generative model. In statistics, this result is known as the Cram\'{e}r-Rao bound of an estimator.  In other words, the well-known bound (upper limit) on the precision of any unbiased estimate of a model parameter in statistics emerges here as a natural consequence of applying information geometry. In the context of the Bayesian brain, this means there is a necessary limit to the certainty with which we can estimate things. We will see next, that attaining this limit translates into attention. 

Notice that the definition of a system immersed in its environment can be extended hierarchically, wherein the gauge theory can be applied at a variety of nested levels. At every step, as the Lagrangian is disturbed (e.g., through changes in forward or bottom-up sensory input), the precision-weighted compensatory forces change to keep the Lagrangian invariant via (backward or top-down) messages. In the setting of predictive coding formulations of variational free energy minimisation, the bottom-up or forward messages are assumed to convey prediction error from a lower hierarchical level to a higher level, while the backward messages comprise predictions of sufficient statistics in the level below. These predictions are produced to explain away prediction errors in the lower level. From the perspective of a gauge theory, one can think of the local forces as prediction errors that increase variational free energy, thereby activating the gauge fields to explain away local forces. 

In this geometrical interpretation, perception and action are educed to form cogent predictions, whereby minimization of prediction errors is an inevitable consequence of the nervous system minimising its Lagrangian. Crucially, the cognitive homologue of precision-weighting is attention, which suggests gauge fields are intimately related to (exogenous) attention. In other words, attention is a force that manifests from the curvature of information geometry, in exactly the same way that gravity is manifest when the space-time continuum is curved by massive bodies. In summary, gauge theoretic arguments suggest that attention (and its neurophysiological underpinnings) constitutes a necessary weighting of prediction errors (or sensory evidence) that arises because the manifolds traced out by the path of least free energy (or least surprise) are inherently curved. 

\section{Discussion}

Complementary to our work in \cite{Sengupta2016}, this paper advances the sketch of an algorithm that utilizes parallel transport for statistical manifolds governed by variational free-energy. It is well known that Riemann conjugate gradient method differs from its Euclidean counterpart (i.e., for small step-sizes where the geometry is close to being Euclidean) only by third order terms \cite{Edelman1998,Bonnabel2013,Raskutti2015} -- enabling these algorithms to converge quadratically near the extremum point. Nevertheless, our algorithm facilitates us to compute discrete approximations of the parallel transport, without requiring us to have any knowledge of the tangent structure of the manifold. This makes it tractable for those manifolds where one need not assume the presence of an ambient space. 

From the point of neuroscience, we consider the principle of free energy minimization as a candidate gauge theory that prescribes neuronal dynamics in terms of a Lagrangian. Here, the Lagrangian is the variational free energy, which is a functional of a probability distribution encoded by neuronal activity. This probabilistic encoding means that neuronal activity can be described by a path or trajectory on a manifold in the space of sufficient statistics (variables that are sufficient to describe a probability distribution). In other words, if one interprets the brain as making inferences, the underlying beliefs must be induced by biophysical representations that play the role of sufficient statistics. This is important because it takes us into the realm of differential geometry (see S2 in \cite{Sengupta2016}), where the metric space -- on which the geometry is defined -- is constituted by sufficient statistics (like the mean and variance of a Gaussian distribution). Crucially, the gauge theoretic perspective provides a rigorous way of measuring distance on a manifold, such that the neuronal dynamics transporting one distribution of neuronal activity to another is given by the shortest path. Such a free energy manifold is curvilinear and finding the shortest path is a non-trivial problem -- a problem that living organisms appear to have solved. It is at this point that the utility of a gauge theoretic approach appears; suggesting particular solutions to the problem of finding the shortest path on curved manifolds. The nature of the solution prescribes a normative theory for self-organized neuronal dynamics. In other words, solving the fundamental problem of minimizing free energy -- in terms of its path integrals -- may illuminate not only how the brain works but may provide efficient schemes in statistics and machine learning

Variational or Monte Carlo formulations of the Bayesian brain require the brain to invert a generative model of the latent (unknown or hidden) causes of sensations (see S3 in \cite{Sengupta2016}). The implicit normative theory means that neuronal activity (and connectivity) maximises Bayesian model evidence or minimises variational free energy (the Lagrangian) -- effectively fitting a generative model to sensory samples. This entails an encoding of beliefs (probability distributions) about the latent causes, in terms of biophysical variables whose trajectories trace out a manifold. In (deterministic) variational schemes, the coordinates on this manifold are the sufficient statistics (like the mean and covariance) of the distribution or belief while for a (stochastic) Monte Carlo formulation, the coordinates are the latent causes themselves. The inevitable habitat of these sufficient statistics (e.g., neuronal activity) is a curved manifold.

This curvature (and associated information geometry) may have profound implications for neuronal dynamics and plasticity. As the Lagrangian is a function of beliefs (probabilities), the manifold that contains this motion is necessarily curved. This means, neuronal dynamics, in a local frame of reference, will (appear to) be subject to forces and drives (i.e., Levi-Civita connections). For example, the motion of synaptic connection strengths (sufficient statistics of the parameters of generative models) depends upon the motion of neural activity (sufficient statistics of beliefs about latent causes), leading to experience-dependent plasticity. A more interesting manifestation may be attention that couples the motion of different neuronal states in a way that depends explicitly on the curvature of the manifold (as measured by Fisher information). 

In brief, a properly formulated gauge theory should, in principle, provide the exact form of neuronal dynamics and plasticity. These forms may reveal the underlying simplicity of many phenomena that we are already familiar with, such as event-related brain responses, associative plasticity, attentional gating, adaptive learning rates and so on.

\begin{appendices}
\section{Software}
\label{appendix:software}

Variational algorithms for (nonlinear) regression, probabilistic graphic models of varying complexity and variational reinforcement learning (active inference; Markov Decision Processes) have been released via the statistical parametric mapping (SPM) toolbox \url{http://www.fil.ion.ucl.ac.uk/spm/}. Figure \ref{fig_s1} provides a screenshot of a wide variety of demos that detail a variety of problems. Along with variational methods the SPM suite also includes stochastic methods (\textit{mci toolbox}) such as Langevin Monte Carlo, Manifold Monte Carlo, Riemannian Markov Chain Monte Carlo (MCMC), Hamiltonian MCMC \cite{Sengupta2015}, population MCMC \cite{Sengupta2015a}, geometric Annealed Importance Sampling algorithms \cite{Penny2016}, amongst others. Parallel transport for variational models shall be made available in subsequent releases.  

\begin{figure}
 \centering \includegraphics[width=0.55\textwidth, height=3in]
 {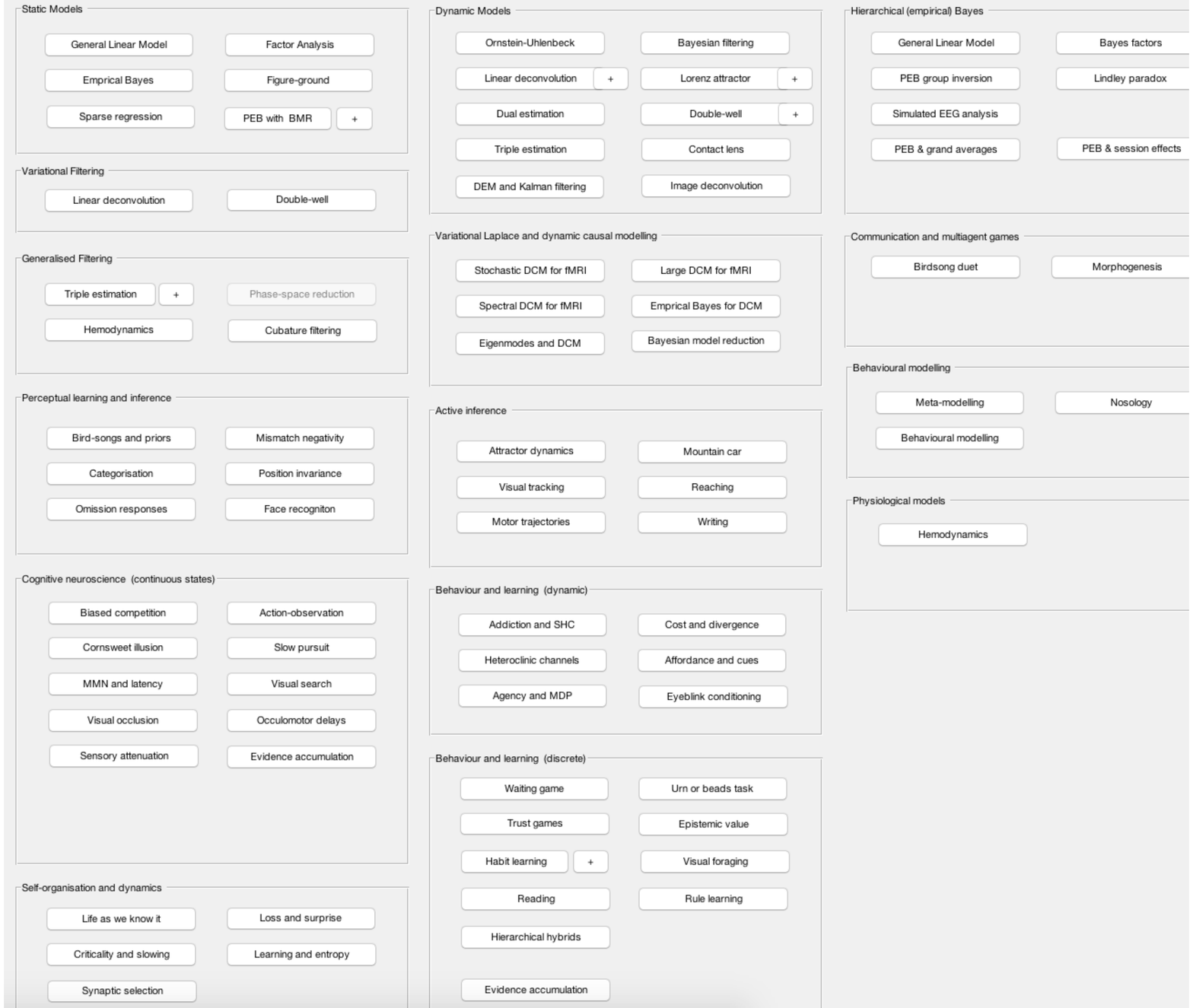}
 \caption{Screenshot of the SPM toolbox that primarily uses variational free energy minimization for a wide-variety of problems in neuroscience, non-linear dynamics, reinforcement learning, amongst others.}
 \label{fig_s1}
\end{figure}

\end{appendices}

{\small
\bibliographystyle{icml2017}
\bibliography{parallel_transport}
}

\end{document}